\documentclass[pra,twocolumn,showpacs,nofootinbib,letterpaper,superscriptaddress]{revtex4}

\usepackage{amsmath} 
\usepackage{graphicx} 
\usepackage{verbatim} 
\usepackage{color} 
\usepackage{subfigure} 
\usepackage{hyperref} 

\begin{document}

\title{Probing dipole-forbidden autoionizing states by isolated attosecond pulses}
\date{\today}
\author{Wei-Chun Chu}
\affiliation{J. R. Macdonald Laboratory, Department of Physics, Kansas State University,
Manhattan, Kansas 66506, USA}
\affiliation{Max Planck Institute for the Science of Light,
G\"{u}nther-Scharowsky-Stra{\ss}e 1, 91058 Erlangen, Germany}
\author{Toru Morishita}
\affiliation{Department of Engineering Science, The University of Electro-communications
1-5-1 Chofu-ga-oka, Chofu-shi, Tokyo 182-8585, Japan}
\author{C.~D.~Lin}
\affiliation{J. R. Macdonald Laboratory, Department of Physics, Kansas State University,
Manhattan, Kansas 66506, USA}

\begin{abstract}
We propose a general technique to retrieve the information of dipole-forbidden resonances in
the autoionizing region. In the simulation, a helium atom is pumped by an isolated attosecond
pulse in the extreme ultraviolet (EUV) combined with a few-femtosecond laser pulse. The excited
wave packet consists of the $^1S$,
$^1P$, and $^1D$ states, including the background continua, near the $2s2p(^1P)$ doubly excited
state. The resultant electron spectra with various laser intensities and time delays between the
EUV and laser pulses are obtained by a multilevel model and an \textit{ab initio} time-dependent
Schr\"{o}dinger equation calculation. By taking the \textit{ab initio} calculation as a
``virtual measurement'', the dipole-forbidden resonances are characterized by the multilevel
model.
We found that in contrast to the common assumption, the nonresonant coupling between the continua
plays a significant role in the time-delayed electron spectra, which shows the correlation effect
between photoelectrons before they leave the core. This technique takes the advantages of
ultrashort pulses uniquely and would be a timely test for the current attosecond technology.
\end{abstract}

\maketitle

\section{Introduction}\label{sec:introduction}

Photoexcitation or photoionization processes are among the most common tools to study the
structures of matters. For atomic or molecular systems, the details in the spectroscopy reveal the
information at the microscopic regime such as the energy levels, decay rates, and transition
strengths between the quantum states. A large amount of data has been obtained and recorded by
synchrotron radiations up to the x-ray regimes. Due to the limited brilliance of the light
sources, or the limited intensity of the fields, normally only single-photon processes are
considered in synchrotron measurements. Although it can be very precise and reliable to
obtain the atomic energy levels, the dipole-forbidden states (DFSs) with respect to the ground
state are not accessible. Nonetheless, these DFSs are essential in numerous optical techniques or
subjects including electromagnetically induced
transparency (EIT)~\cite{harris, fleischhauer}, two-photon absorption~\cite{abella, rumi}, Raman
spectroscopy and more. It is noteworthy that some of these or similar quantum phenomena
have been reconsidered in ultrafast optical configurations~\cite{linmf, chen, loh, ott1, ott2} and
attract much attention, where their interpretations are based on the dynamics involving DFSs.
Until now the aspiration for better measurements or characterizations of DFSs is tremendous.

In recent years, there are renewed interests in the studies of Fano resonances and the
corresponding autoionizing states (AISs), which have long been perceived in the energy domain as
interpreted by Fano more than 50 years ago~\cite{fano}. The new interests are founded on the
developments in ultrashort light sources and ultrafast optical techniques. The
EIT effect between the coupled AISs was observed by time-delayed transient absorption in the
femtosecond timescale~\cite{loh}. The time-domain measurements of the autoionization dynamics using
attosecond pulses were reported in argon~\cite{wang} and in helium~\cite{gilbertson}. Most
recently, the Fano line shapes modulated by time-delayed intense fields in transient
absorption spectroscopy have been measured and analyzed~\cite{ott1, ott2}, which has also brought
to light the astounding potential to shape and to control the attosecond pulses~\cite{chujpb}.
In these studies and the related theoretical works, the dynamics of the Fano wave packet were
mostly implemented as attosecond streaking~\cite{wickenhauser, zhaozx} or resonant
coupling~\cite{chu11, chu12, pfeiffer, chu13} of the Fano states, if not carried out by
\textit{ab initio} calculations~\cite{zhaoj, tarana, argenti} which were numerically expensive
and less intuitive. While the attention has been put
on the directly measurable resonances under the dressing field, the properties of the
dipole-forbidden resonances (DFRs) in the coupling were largely overlooked. In this work, we aim
to provide a model describing the realistic coupling between the AISs where the DFRs can be
``probed'', and their essential properties can be characterized.

Taking the helium atom in the energy range near the $2s2p(^1P)$ resonance as an example, we carry
out an \textit{ab initio} two-active-electron time-dependent Schr\"{o}dinger equation
(TAE-TDSE) based on the time-dependent hyperspherical (TDHS) method~\cite{morishita, hishikawa,
liu} to virtually provide the reference experimental data. Then a multilevel model with
parametrized atomic structure is developed and applied to the case, where the comparison to the
virtual experiment optimizes the atomic parameters of the DFRs. In our previous model consisting
of coupled AISs~\cite{chu13}, the transitions between the AISs were reduced to the transitions
involving only the bound states--bound-bound (B-B) transitions--by ignoring the second-order
(two-electron) transitions. This naive picture successfully predicted the inversion
of Fano line shapes under intense dressing fields, but the effects on the DFRs were missing. The
present model includes the transitions between the discrete states and the background continua
(bound-continuum, B-C), and between the background continua of different symmetries
(continuum-continuum, C-C). With ultrashort dressing fields, the significance of the B-C and C-C
transitions is justified because these transitions take place in such a short time when electrons
are still
near the ionic core, and the distorted continuum waves are far from plane waves. This improved
picture emphasizes the influences of these added transitions in the dressing field on the Fano
line shapes of all resonances, including the DFRs. This signature broadly implies that in
ultrafast dynamics, even in weak fields in the sense of few-photon transitions, the implementation
or neglection of the C-C transitions needs to be reconsidered carefully, especially for
ultrashort overlapping pulses.

The structure of this article is in the following. Section~\ref{sec:model} gives detail accounts
of the formulation of the present model for the multilevel autoionizing systems and its differences
from previous models. The limitation of the current model is discussed.
Section~\ref{sec:measurement} connects the laser-dressed angle-resolved electron spectroscopy--the
preassumed measurement--to the retrieval of the parameters in the model. Section~\ref{sec:ml-tdse}
compares the calculations by the present model and by TAE-TDSE to determine the atomic parameters.
Using the present model with the retrieved parameters, Sec.~\ref{sec:time-delayed} presents the
time-delayed electron spectra by the synchronized attosecond extreme ultraviolet (EUV) and strong
laser pulses, as such a scheme is commonly performed experimentally to extract the ultrafast
dynamics. The role of the C-C coupling in the
model is elaborated in Sec.~\ref{sec:c-c}. Finally Sec.~\ref{sec:conclusions} gives the
concluding remarks. In this paper, we use eletron volts (eV) and femtoseconds (fs) except in
Sec.~\ref{sec:model}, unless otherwise specified. Field intensity is defined by the
cycle-averaged values.

\section{Model}\label{sec:model}

The present model describes the dynamics of an autoionizing wave packet in an atom consisting of
the AISs of two different symmetries, which are pumped by a weak EUV attosecond pulse and dressed
by a relatively strong (TW/cm$^2$) femtosecond laser pulse. The goal is to calculate the
photoelectron spectrum with a given set of fields and atomic parameters. Providing the angular
distribution of the electrons in real measurements, we assume that the spectrum of each spherical
partial wave can be isolated. The separation of partial waves in the experimental point of view
will be further discussed in Sec.~\ref{sec:results}. Each partial wave is then compared with the
result from the TAE-TDSE calculation, which is described in details in Refs.~\cite{hishikawa, liu}.
By optimizing the modeled spectra, the atomic parameters are retrieved. In this section, atomic
units (a.u.) are used unless otherwise specified.

\begin{figure}[htbp]
\centering
\includegraphics[width=0.47\textwidth]{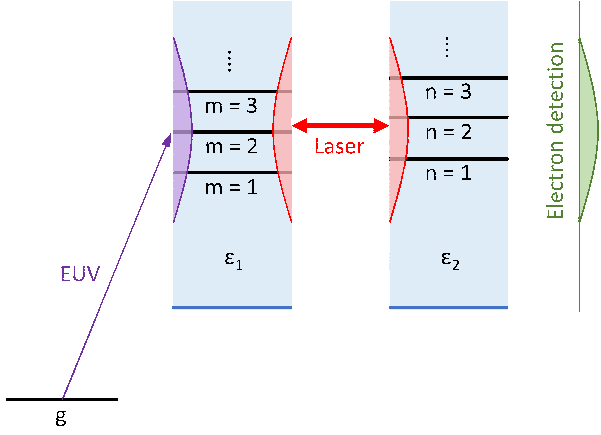}
\caption{(Color online) schematics of the coupled autoionizing system. The purple and red arrows
are the transitions by the EUV and laser pulses. The bandwidths of the EUV covers a group of AISs
$|m\rangle$ and the background continuum $|\epsilon_1\rangle$. The laser couples $|m\rangle$ and
$|\epsilon_1\rangle$ to states of other symmetries by single-photon transitions following the
selection rule. The final wave packet after the pulse is projected out in the photoelectron
spectroscopy.}
\label{fig:scheme}
\end{figure}

The autoionizing system pumped by the EUV and strongly coupled by the laser is schematically
plotted in Fig.~\ref{fig:scheme}. The total wave function of the atomic system is written as
\begin{align}
|\Psi(t)\rangle &= e^{-i\epsilon_gt} c_g(t) |g\rangle \notag\\
&+ e^{-i\epsilon_et} \left[ \sum_m{c_m(t)|m\rangle} + \int{ c_{\epsilon_1}(t) |\epsilon_1\rangle d\epsilon_1} \right.\notag\\
& \hspace{1.2 cm} \left. + \sum_n{c_n(t)|n\rangle} + \int{ c_{\epsilon_2}(t) |\epsilon_2\rangle d\epsilon_2} \right],
\label{eq:Psi}
\end{align}
where $|g\rangle$ is the ground state with energy $\epsilon_g$, and $\epsilon_e \equiv
\epsilon_g+\omega_X$ corresponds to the central energy level pumped by the EUV, where $\omega_X$
is the photon energy of the EUV. The fast oscillation in terms of the EUV frequency is factored
out, where the $c(t)$
coefficients are slowly varying in time. The bound excited states $|m\rangle$ and $|n\rangle$ are
embedded in the background continua $|\epsilon_1\rangle$ and $|\epsilon_2\rangle$, respectively,
where the labels $|m\rangle$ and $|\epsilon_1\rangle$ are used exclusively for dipole-allowed
states, and $|n\rangle$ and $|\epsilon_2\rangle$ are used exclusively for dipole-forbidden states.
The Hamiltonian of the system is $H(t)=H_A-\left[E_X(t)+E_L(t)\right]D$, where $H_A$ is the atomic
Hamiltonian, $E_X(t)$ and $E_L(t)$ are the electric fields of the EUV and the laser, respectively,
and $D$ is the dipole operator. Thus, in this basis set, the off-diagonal elements in the
Hamiltonian are
\begin{align}
\langle g|H|m \rangle &= -D_{gm} E_X(t), \\
\langle g|H|\epsilon_1 \rangle &= -D_{g\epsilon_1} E_X(t), \\
\langle m|H|n \rangle &= -D_{mn} E_L(t), \\
\langle m|H|\epsilon_2 \rangle &= -D_{m\epsilon_2} E_L(t), \\
\langle \epsilon_1|H|n \rangle &= -D_{\epsilon_1n} E_L(t), \\
\langle \epsilon_1|H|\epsilon_2 \rangle &= -D_{\epsilon_1 \epsilon_2} E_L(t),
\end{align}
for the dipole transitions, and
\begin{align}
\langle m|H|\epsilon_1 \rangle &= V_{m\epsilon_1}, \\
\langle n|H|\epsilon_2 \rangle &= V_{n\epsilon_2},
\end{align}
for the configuration interactions responsible for autoionization, 
and $D$ and $V$ are
determined by the atomic structure. Note that the dimensions of the Hamiltonian matrix elements
are $\epsilon$, $\sqrt{\epsilon}$, and 1 for B-B, B-C, and C-C transitions, respectively,
where $\epsilon$ denotes energy. For simplicity, we employ the standing wave representation for
the continuum states in Eq.~(\ref{eq:Psi}), so all $D$ and $V$ are real quantities.

Certain approximations are adopted to simplify the model. The type of measurement in our concern
involves a perturbative EUV pulse with the bandwidth of a few eVs, and a few-cycle laser pulse in
the infrared to visible energy region and with the intensity up to the TW/cm$^2$ scale.
The EUV field is written in the form of
\begin{equation}
E_X(t) = F_X(t) e^{i\omega_Xt} + F_X^*(t) e^{-i\omega_Xt}, \label{eq:FX}
\end{equation}
where $F_X(t)$ is the field envelope containing any additional phase other than the carrier
phase. The rotating wave approximation (RWA) can be applied to the EUV, where for each transition,
only one of the two terms in Eq.~(\ref{eq:FX}) is taken into account. For the laser pulse, RWA
is not suitable since the bandwidth for a few-cycle pulse is comparable to its central frequency.
The wave packet is in the energy range much higher than the binding energy so that the background
continua change only slightly across each resonance, where the $D$ and $V$ elements are constants
of energy estimated at each resonance. In the following derivation, the energy dependence of
$D$ and $V$ is kept in the expression for clarity if necessary, but is removed in the numerical
evaluation.

By solving TDSE with the total wave function in Eq.~(\ref{eq:Psi}), the equation of motion (EOM),
or the coupled equations for all the $c(t)$ coefficients can be established. By applying adiabatic
eliminations (AE) of the continua, i.e., assuming $\dot{c}_{\epsilon_1}(t) =
\dot{c}_{\epsilon_2}(t) = 0$ in the original coupled equations for the continuum, the preliminary
forms of the continuum-state coefficients are obtained as
\begin{align}
c_{\epsilon_1}^p(t) =& \frac{1}{\delta_{\epsilon_1}} \Bigg[ -F_X^*(t) D_{\epsilon_1 g} c_g(t) + \sum_m V_{\epsilon_1 m} c_m(t) \notag\\
& - E_L(t) \sum_n D_{\epsilon_1 n} c_n(t) \Bigg], \\
c_{\epsilon_2}^p(t) =& \frac{1}{\delta_{\epsilon_2}} \Bigg[ \sum_n V_{\epsilon_2 n} c_n(t) - E_L(t) \sum_m D_{\epsilon_2 m} c_m(t) \Bigg],
\end{align}
where $\delta_1 \equiv \epsilon_e - \epsilon_1$ and $\delta_2 \equiv \epsilon_e - \epsilon_2$ are
the detuning of the continuum states. With the assumption $c_g(t)=1$ for the perturbative EUV,
the preliminary forms are then inserted back to the EOM, which gives the coefficients of the bound
excited states by
\begin{align}
i\dot{c}_m(t) =& -F_X^*(t) \bar{D}_{mg} - (\delta_m+i\kappa_m)c_m(t) \notag\\
&- E_L(t) \sum_n{\bar{D}_{mn}c_n(t)}, \label{eq:cm}\\
i\dot{c}_n(t) =& -iF_X^*(t)E_L(t)j_{ng} - (\delta_n+i\kappa_n)c_n(t) \notag\\
&- E_L(t) \sum_m{\bar{D}_{nm}c_m(t)}, \label{eq:cn}
\end{align}
where $\delta_m \equiv \epsilon_e-\epsilon_m$ and $\delta_n \equiv \epsilon_e-\epsilon_n$ are the
detuning of the AISs, $\kappa_m \equiv \Gamma_m/2 = \pi |V_{m\epsilon_1}|^2$ and $\kappa_n \equiv
\Gamma_n/2 = \pi |V_{n\epsilon_2}|^2$ are the half widths of the resonances, and $j_{gn} \equiv
\pi D_{g\epsilon_1}D_{\epsilon_1n}$ is the light-induced broadening of the DFRs. The
composite dipole matrix elements $\bar{D}$ are defined by
\begin{align}
\bar{D}_{gm} &\equiv D_{gm} - i\pi D_{g\epsilon_1}V_{\epsilon_1m}, \label{eq:Hgm} \\
\bar{D}_{mn} &\equiv D_{mn} - \pi^2 V_{m\epsilon_1}D_{\epsilon_1 \epsilon_2}V_{\epsilon_2n} \notag\\
& -i\pi (V_{m\epsilon_1}D_{\epsilon_1n} + D_{m\epsilon_2}V_{\epsilon_2n}) \label{eq:Hmn}
\end{align}
The elements $\bar{D}_{gm}$ and $\bar{D}_{mn}$ are responsible for the EUV and laser transitions,
respectively. For a given $\bar{D}_{gm}$, the Fano $q$ parameter of the $|m\rangle$ state is
easily identified as
\begin{equation}
q_m = \frac{D_{gm}}{\pi D_{g\epsilon_1} V_{\epsilon_1m}}. \label{eq:qm}
\end{equation}
For each dipole-allowed state $|m\rangle$, $q_m$ only takes the EUV transition from the ground
state into account, and is the commonly known and measured Fano line shape parameter.
On the other hand, for each DFR $|n\rangle$, the $q$ parameter needs to consider the laser
transitions from all possible paths, and to represent the phase induced by the interference
between the autoionization paths and the direct ionization paths. Each $\bar{D}_{mn}$ counts only
one pair of $|m\rangle$ and $|n\rangle$ states, and the general line shapes of $|n\rangle$ is
determined by the collection of $\bar{D}_{mn}$ of all $m$ under the laser bandwidth.
As one of
the approximations, because the structures of the background continua change sufficiently slowly
in the energy ranges in concern, $\Gamma_m$, $\Gamma_n$, $\bar{D}_{gm}$, $\bar{D}_{mn}$, $q_m$,
and $j_{gn}$ are constant of energy estimated just at the single energy points
$\epsilon_1=\epsilon_m$ and $\epsilon_2=\epsilon_n$, except that $j_{gn}$ is estimated at
$\epsilon_1=\epsilon_e$.

After the calculation of the bound-state coefficients $c_m(t)$ and $c_n(t)$, we aim to recover
the continuum-state coefficients $c_{\epsilon_1}(t)$ and $c_{\epsilon_2}(t)$ from the preliminary
forms. Note that AE works for the treatment in Eqs.~(\ref{eq:cm}) and (\ref{eq:cn}) because only
thecollective effects by the continua are needed there, which are not strongly affected by the
detail shapes of the continua. Here, when the goal is to acquire the electron spectra, we have to
go beyond AE and keep the previously eliminated terms, i.e.,
$i\dot{c}_{\epsilon_1}(t)/\delta_{\epsilon_1}$ and $i\dot{c}_{\epsilon_2}(t)/\delta_{\epsilon_2}$,
in $c_{\epsilon_1}^p(t)$ and $c_{\epsilon_2}^p(t)$, respectively. The corrected preliminary forms
are then employed in the mutual coupling terms between the continua to finally obtain
\begin{align}
i\dot{c}_{\epsilon_1}(t) =& -F_X^*(t)D_{\epsilon_1 g} -\delta_{\epsilon_1}c_{\epsilon_1}(t) +\sum_m{V_{\epsilon_1m} c_m(t)} \notag\\
& -E_L(t) \left[ \sum_n{\bar{D}_{\epsilon_1n}c_n(t)} +\alpha_{\epsilon_1}(t) \right], \label{eq:c1}\\
i\dot{c}_{\epsilon_2}(t) =& -iF_X^*(t)E_L(t) j_{\epsilon_2 g} - \delta_{\epsilon_2} c_{\epsilon_2}(t) + \sum_n{V_{\epsilon_2n} c_n(t)} \notag\\
& - E_L(t)\left[ \sum_m{\bar{D}_{\epsilon_2m}c_m(t)} +\alpha_{\epsilon_2}(t) \right], \label{eq:c2}
\end{align}
where the composite dipole matrix elements therein are defined by
\begin{align}
\bar{D}_{\epsilon_1n} &\equiv D_{\epsilon_1n} - i\pi D_{\epsilon_1\epsilon_2}V_{\epsilon_2n}, \label{eq:H1n}\\
\bar{D}_{m\epsilon_2} &\equiv D_{m\epsilon_2} - i\pi V_{m\epsilon_1}D_{\epsilon_1\epsilon_2}. \label{eq:Hm2}
\end{align}
The direct coupling terms between the continua (not via any resonances), given by
\begin{align}
\alpha_{\epsilon_1}(t) &\equiv \left.\pi D_{\epsilon_1\epsilon_2} \dot{c}_{\epsilon_2}(t) \right|_{\epsilon_2=\epsilon_e}, \label{eq:alpha1} \\
\alpha_{\epsilon_2}(t) &\equiv \left.\pi D_{\epsilon_1\epsilon_2} \dot{c}_{\epsilon_1}(t) \right|_{\epsilon_1=\epsilon_e}, \label{eq:alpha2}
\end{align}
are calculated in each time step by Eqs.~(\ref{eq:c1} and \ref{eq:c2}) without the $\alpha(t)$
terms, and they are then added back as iterative corrections. The coupling between the continua
through
$\alpha(t)$ greatly reduce the number of coupled equations in the numerical calculation from
$2N^2$ to $2N$, where $N$ is the number of energy steps in the range in concern. An 8-eV wide
wave packet with the energy resolution of 20~meV will require $N=400$, which is typical for an
IAP+IR excitation. The validity of the $\alpha(t)$ terms relies on the smoothness of
the continuum coefficients, which makes possible a simple collective effect for the other
continuum in the coupling.

By Eqs.~(\ref{eq:cm}), (\ref{eq:cn}), (\ref{eq:c1}), and (\ref{eq:c2}), and with a given set of
atomic and field parameters, the total wave function in the form of Eq.~(\ref{eq:Psi}) is uniquely
determined. The electron spectra for the two symmetries are $P_1(\epsilon_1) =
|c_{\epsilon_1}(t_f)|^2$ and $P_2(\epsilon_2) = |c_{\epsilon_2}(t_f)|^2$ where $t_f$ is much
larger than all the resonance lifetimes after the coefficients are stablized~\cite{chu11}.
Alternative to Eq.~(\ref{eq:Psi}), the total wave function can also be written in atomic
eigenstates, where the coefficients $\bar{c}_{\epsilon_1}(t_f)$ and $\bar{c}_{\epsilon_2}(t_f)$,
associated with the two symmetries, respectively, are stablized right after the end of the
external field. Thus, in cases with ultrashort pulses, employing the form of the total wave
function in eigenstate basis cuts down the calculation time significantly. The conversion between
the eigenstate coefficients and the bound-state and continuum-state coefficients has been detailed
in Ref.~\cite{chu11}, and is given by
\begin{equation}
\bar{c}_\epsilon(t) = \left( i\sin \theta_\epsilon - \cos \theta_\epsilon \right) c_\epsilon(t)
+ \cos \theta_\epsilon \sum_l{\frac{\tan \theta_{\epsilon l}}{\pi V_l} c_l(t)}, \label{eq:cE}
\end{equation}
where
\begin{equation}
\tan \theta_{\epsilon l} \equiv -\frac{\kappa_l}{\epsilon-\epsilon_l},
\end{equation}
and
\begin{equation}
\tan \theta_\epsilon \equiv \sum_l \tan \theta_{\epsilon l},
\end{equation}
where $\epsilon$ and $l$ are the indices belonging to the same symmetry ($\epsilon_1$, $m$ or
$\epsilon_2$, $n$ in the present case).

The inclusion of the laser coupling involving continuum states, particularly the C-C
transitions that have been ignored previously, raises the dressing intensities that could be
handled properly. Nonetheless, we should stress the restrictions of this model in the following.
First of all, the model works only in the presence of bound states on both sides of the coupling.
The assumption that the continua evolve more slowly than the bound states is required for AE,
which relies on the existence of near-resonant bound states to mediate all transitions. Second,
the description of the system is in terms of atomic angular momentum eigenstates, which is not
suitable for the strong field ionization in the streaking regime~\cite{itatani}. In such a regime,
the momentum of emitted electrons are shifted in the direction parallel to the polarization of
field, so that the wave packet in general consists of a large number of angular momenta.

\section{Results and analysis}\label{sec:results}

\subsection{Partial wave analysis of photoelectrons}\label{sec:measurement}

We apply the general technique described above to the simplest autoionizing system, helium atom,
in the energy range centered at the lowest dipole-allowed doubly excited state $2s2p(^1P)$
(referred by $2s2p$ hereafter). With linearly polarized EUV and laser pulses and following the
selection rules, the final detected photoelectron wave packet is composed by $^1S^e$, $^1P^o$, and
$^1D^e$ states of magnetic quantum number $m=0$, under the constraint that the laser intensity is
within the limit of 1~TW/cm$^2$ so that other symmetries are negligible. Thus following the same
derivation introduced in the model, the third symmetry is included in the coupled equations, and
its coefficients are solved accordingly.

The excited wave packet $|\Psi_E(t)\rangle$, consisting of the excited bound and continuum states
as defined in Eq.~(\ref{eq:Psi}), is written as
\begin{equation}
|\Psi_E(t)\rangle = e^{-i\epsilon_et} \sum_{l=0}^2 {\int{ \bar{c}_{\epsilon l}(t) |\epsilon l\rangle d\epsilon }}, \label{eq:PsiE}
\end{equation}
where $|\epsilon l\rangle$ are the atomic eigenstates with electron energy $\epsilon$ and angular
momentum $l$, following the conventional atomic orbital notations. Note that in this expression,
the core orbital is skimmed because in our case and the energy range in concern, the core orbital
is always $1s$ (the excitation energy of $2s$ and $2p$ from the ionic ground state is 40.8~eV,
which is 5.3~eV higher
than the $2s2p$ level). The angular momentum of the photoelectron in the $s$, $p$, or $d$
partial wave is already the total angular momentum. The symmetry in the model description is
directly mapped onto the angular momentum of the photoelectron in the present calculation. The
shorthands $S$, $P$, and $D$ are used hereafter for the three symmetries of the total wave
function in concern.

For detecting photoelectrons with angular distribution, the wave packet in Eq.~(\ref{eq:PsiE})
should be projected onto the final state of momentum $\vec{k}$ that satisfies incoming boundary
conditions and represents the ejected photoelectron~\cite{bransden}.
In particular, it is written in the energy-normalized form by
\begin{equation}
\psi^{(-)}_{\vec{k}}(\vec{r}) = \sqrt{\frac{2}{\pi k}} \frac{1}{r} \sum_{lm} i^l e^{-i \eta_l} {u_l(kr) Y_l^m(\hat{r}) Y_l^{m*}(\hat{k})}, \label{eq:psik}
\end{equation}
where $\eta_l$ are the total phase shift due to the atomic potential and $u_l(kr)$ are the radial
waves corresponding to $|\epsilon l\rangle$.
The measured signal is
$P(\vec{k}) = |\langle \psi^{(-)}_{\vec{k}} |\Psi_E(t_f) \rangle|^2$, where
$t_f$ is the detection time, which is infinite after the ionization in the atomic timescale. 
The angle- and energy-resolved electron yield, corresponding to the $\vec{k}$ component of the
wave packet, is thus
\begin{align}
P(\epsilon,\theta) &= |\langle \psi^{(-)}_{\vec{k}} |\Psi_E(t_f) \rangle|^2 \notag \\
&= \left| \sum_{l=0}^2 \sqrt{\frac{2l+1}{4\pi}} \frac{e^{i\eta_l}}{i^l} {\bar{c}_{\epsilon l}(t_f) P_l(\cos \theta)} \right|^2, \label{eq:Pk}
\end{align}
where $\epsilon = k^2/2$, $\theta$ is the polar angle of the momentum direction $\hat{k}$ with
respect to the polarization of fields, and the azimuth dependence has been removed by considering
only $m=0$.

In the above, we have described how the electron yield $P(\epsilon,\theta)$ is related to the final
wave packet of each partial wave expressed by $e^{i\eta_l}\bar{c}_{\epsilon l}(t_f)$. Nonetheless,
with $P(\epsilon,\theta)$ given by a photoelectron measurement, the process to assign
$e^{i\eta_l}\bar{c}_{\epsilon l}(t_f)$ should be achieved by numerical fitting with caution. In
the current case, for each energy point, the three complex coefficients corresponding to $l=0$,
1, 2 are fitted for the signal intensity as a function of $\theta$. The energy range in concern is
far above the binding energy, and the energy dependence of the phase of
$e^{i\eta_l}\bar{c}_{\epsilon l}(t_f)$ is significant only across the resonances. In our case, the
phase of $e^{i\eta_2}\bar{c}_{\epsilon d}(t_f)$ is set as an arbitrary constant of
$\epsilon$ since there are no $l=2$ resonances in this energy range. In this way, the coefficients
of all partial waves in the whole energy region in concern are extracted from the experiment, i.e.
we convert $P(\epsilon,\theta)$ to $e^{i\eta_l}\bar{c}_{\epsilon l}(t_f)$ as the main experimental
data to be compared with.

\subsection{Retrieval of atomic structure parameters}\label{sec:ml-tdse}

\begin{figure}[htbp]
\centering
\includegraphics[width=0.50\textwidth]{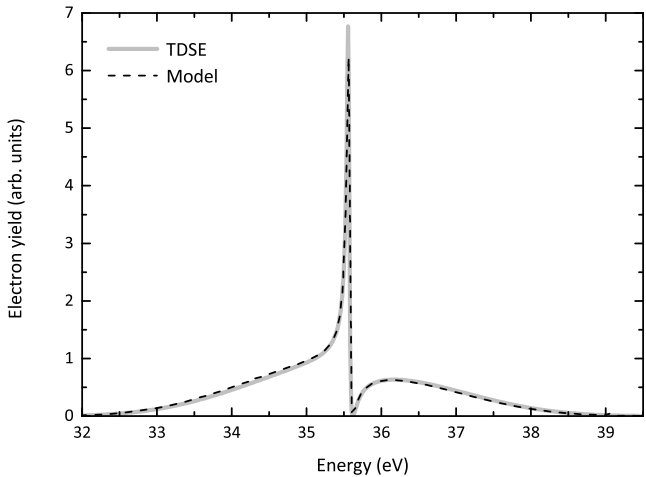}
\caption{Photoionization electron spectra near the He $2s2p$ resonance by a 690-as EUV pulse,
calculated by TAE-TDSE and by the multilevel model. The EUV pump is weak, and only the $p$ partial
wave is nonnegligible. The energy, width, and $q$ parameter of the resonance are retrieved by
fitting with the TAE-TDSE.}
\label{fig:x}
\end{figure}

In the following, the TAE-TDSE calculation is taken as a virtual experiment, and the parameters
in the model calculation are adjusted to achieve the best agreement. Since TAE-TDSE calculates
photoelectron spectra in partial waves, the step to ``dissemble'' the measured signal into
partial waves described in Sec.~\ref{sec:measurement} is skipped in our comparison to TAE-TDSE.
In the physical process, a weak broadband EUV pulse is applied first to ionize the system. The
duration is 690~as and the peak intensity
is $10^9$~W/cm$^2$. It excites a wave packet consisting of the $2s2p$ AIS and the
$1s\epsilon p$ continuum about 4~eV wide around the resonance. In this laser-free condition, any
states with symmetries other than $P$ are negligible. The TAE-TDSE result of the electron spectrum
is shown in Fig.~\ref{fig:x} by the gray solid curve. The $2s3p(^1P)$ resonance appearing at
39.1~eV is very weakly pumped and neglected in this study.
With only the TAE-TDSE result, the Fano parameters can already be characterized by the Fano
line shape formula~\cite{fano}
\begin{equation}
\sigma(\epsilon) = \sigma_0 \frac{(q+\epsilon)^2}{1+\epsilon^2}, \label{eq:fano}
\end{equation}
combined with the bandwidth profile of pulse, where $\epsilon \equiv 2(E-E_r)/\Gamma$ is the
photon energy relative to the resonance energy $E_r$ and normalized by the width $\Gamma$. Thus
$\Gamma=41$~meV and the line shape $q=-2.66$ for the $2s2p$ resonance are extracted.
The obtained parameters agree well with the actual experimental values $\Gamma=37$~meV and
$q=-2.75$~\cite{domke}. Note that $E_r$ in the model can be arbitrarily shifted by the binding
energy and is not uniquely defined. With such parameters, the multilevel model reproduces the
line shape very
well. By fitting the overall signal strength with TAE-TDSE, the absolute values of the dipole
matrix elements $D_{1s^2,2s2p}=0.036$~a.u. and $D_{1s^2,\epsilon_p}=-0.278$~a.u. are also
obtained, where the ratio between them is already determined by $q$. This retrieval process
relies on the frequency profile of the EUV pulse, which should
be considerably wider than the resonance width and very smooth in frequency. A high-quality
isolated attosecond pulse could do the job well.

\begin{figure}[htbp]
\centering
\includegraphics[width=0.50\textwidth]{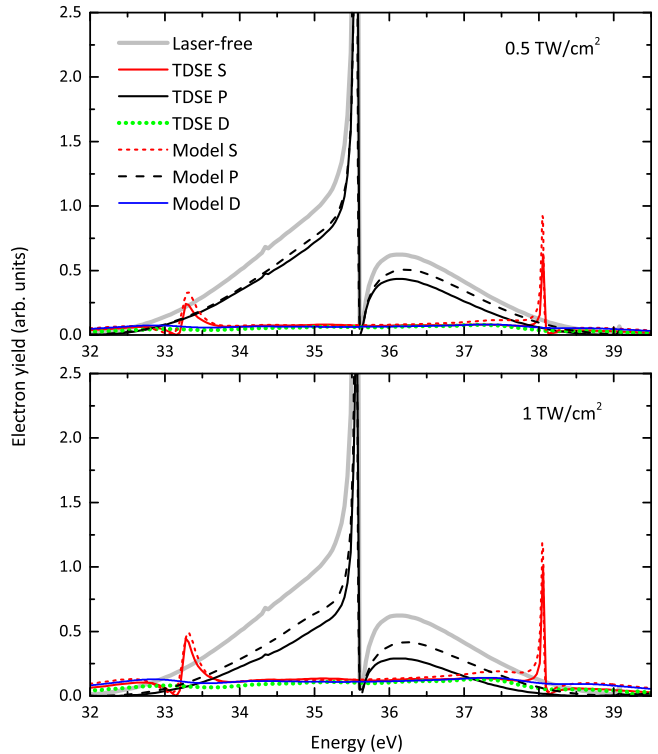}
\caption{(Color online) Laser-dressed EUV electron spectra of the dominant partial waves for the
peak laser intensities of 0.5 and 1~TW/cm$^2$, calculated by TAE-TDSE and the multilevel model.
For such moderately intense dressing fields, $\epsilon p$ still surpasses while $\epsilon s$ and
$\epsilon d$ increase with the intensity of the dressing field. The atomic
parameters in the model calculation are optimized for best agreement in the $I_0=0.5$~TW/cm$^2$
case. When the laser intensifies, more signals are ``lost'' due to the ionization by the
laser, which is not counted by the multilevel model and responsible
for the descrepancies in the comparison to the TAE-TDSE results.}
\label{fig:xl}
\end{figure}

In the presence of the laser pulse, the wave packet promoted by the EUV is distorted and projected
onto a combination of symmetries of bound and continuum states. Nevertheless, in the intermediate
intense laser, $P$ is still the dominant symmetry, followed by $S$ and $D$. For overlapping EUV
and laser pulses, i.e., in the time delay $t_0=0$ condition ($t_0$ is defined by the temporal shift
between the two pulse peaks and is positive when the EUV comes first), the electron spectra by
TAE-TDSE of the dominant partial waves are shown in Fig.~\ref{fig:xl}. The laser pulse is 4-fs
(FWHM) long with the central wavelength of 540~nm, which corresponds to the central frequency of
$\omega_L=2.3$~eV. The carrier-envelope phase is set to 0, which
means that at $t_0=0$, the EUV is at the maximum of the laser field in the positive polarization
direction. The top and bottom panels of the figure are for peak laser intensities $I_0=0.5$
and 1~TW/cm$^2$, respectively. Two resonances clearly show up in $\epsilon s$ at about 33.3 and
38.1~eV while $\epsilon d$ is spectrally flat. These resonances have been reported in earlier
works~\cite{oza, ho, lindroth} and are identified as $2s^2(^1S)$ and $2p^2(^1S)$ (referred by
$2s^2$ and $2p^2$ hereafter), respectively. We note that $2p^2(^1D)$ at 35.3~eV~\cite{lindroth} is
not excited by the laser pulse considered here, so it is not taken into account in the following
analysis.

To retrieve the parameters of the dipole-forbidden states, we take the $I_0=0.5$~W/cm$^2$ case in
Fig.~\ref{fig:xl} as the reference. In the region between 34 and 37~eV where no resonances are
involved, the C-C dipole matrix elements $D_{\epsilon p,\epsilon s}$ and $D_{\epsilon p,\epsilon
d}$ control the absolute signal strengths of $\epsilon s$ and $\epsilon d$ exclusively.
The fitting gives $D_{\epsilon p,\epsilon s}=29$~a.u. and $D_{\epsilon p,\epsilon d}=27$~a.u. For
the resonances $2s^2$ and $2p^2$, three parameters for each resonance are extracted at the same
time by fitting its line shape, which are $\Gamma_{2s^2}=225$~meV, $D_{2s2p,2s^2}=1.5$~a.u., and
$D_{1s\epsilon p,2s^2}=2$~a.u. for $2s^2$ and $\Gamma_{2p^2}=35$~meV, $D_{2s2p,2p^2}=1.8$~a.u.,
and $D_{1s\epsilon p,2p^2}=2$~a.u. for $2p^2$. The signals surrounding $2s^2$ and $2p^2$--most
significantly at 37.5 and 38.5~eV--form small bumps away from the resonances, which further
determine $D_{2s2p,1s\epsilon s}=1$~a.u. The $\epsilon d$ spectrum is quite insensitive to
$D_{2s2p,1s\epsilon d}$ which can be set as 0. Now all the atomic structure parameters have been
obtained. We keep the same parameters for further simulations with arbitrary dressing
fields. As seen in the bottom panel of Fig.~\ref{fig:xl}, as the laser intensifies, the $\epsilon
p$ spectrum decreases, and the $\epsilon s$ and $\epsilon d$ spectra, including the resonance
part and the background part, increase.
This means that the laser transfers the electrons from the $2s2p$ resonance--including its
background continuum--to the $S$ and $D$ states, mainly by single-photon transitions.

With the same set of parameters, which have been retrieved in the $I_0=0.5$~TW/cm$^2$ case, the
discrepancy between the multilevel model and TAE-TDSE grows with the dressing field intensity,
as seen in Fig.~\ref{fig:xl}. It is especially obvious in the depletion of the $\epsilon p$
spectrum. This discrepancy comes from the laser ionization of the wave packet, which is not
considered by our model. For the ionization from $2s2p$ to $2l \epsilon l'$, with the binding
energy $I_p=5.3$~eV, the
central laser frequency $\omega_L=2.3$~eV, and the peak laser intensity $I_0=0.5$~TW/cm$^2$, the
Keldysh parameter ($\gamma = \sqrt{2I_p} \omega_L /E_0$ where $E_0$ is the peak field strength) is
13.8. This means that multiphoton ionization prevails over tunnel ionization. Considering the
binding energies of $2s2p$, $2s^2$, and $2p^2$, their multiphoton ionization processes are at least
third-order. Thus with the present laser intensities, the multiphoton ionization should not
overpower the coupling between the resonances, and will not change the conclusion of this
technique. The precise calculation of ionization rate of doubly excited states is a research field
of its own and beyond the scope of this report.

\subsection{EUV-plus-VIS spectroscopy}\label{sec:time-delayed}

\begin{figure}[htbp]
\centering
\includegraphics[width=0.50\textwidth]{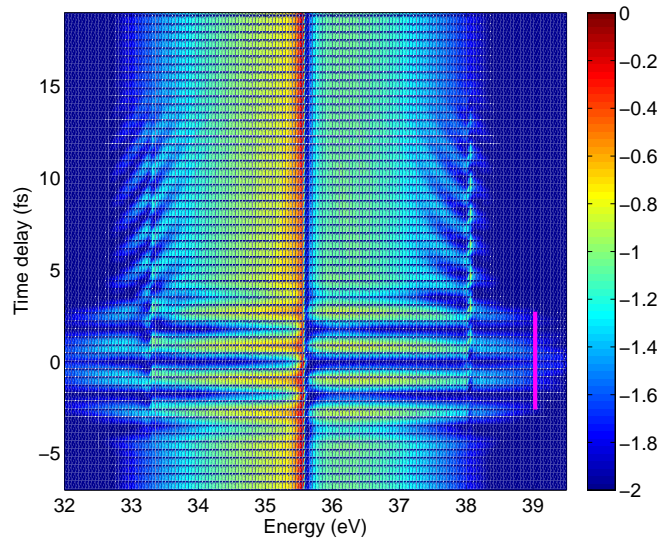}
\caption{(Color online) Angular differential electron spectra in the forward direction along the
polarization by the model calculations. The signals are normalized and plotted in log scale
(color bar is labeled by $\log_{10} \sigma$ where $\sigma$ is the signal strength) with
20~dB dynamic range. The same parameters in Fig.~\ref{fig:xl} are applied, but with fixed laser
intensity ($I_0=0.5$~TW/cm$^2$). The fringes near the overlap of the pulses form streakinglike
patterns. The signals corresponding to the $2s^2$ and $2p^2$ resonances form sidebandlike
structures. The magenta bar on the right indicates the pulse length (FWHM) of the dressing laser.}
\label{fig:delay_0}
\end{figure}

To study the dynamic optical response of the system, instead of tuning the laser intensity,
experimentally it is easier to scan the time delay between the pulses while keeping the intensity
the same. With the fixed $I_0=0.5$~TW/cm$^2$ and the detector direction at 0 degree (along the
positive polarization direction of light), the angular-differential time-delay electron spectra
by the multilevel model are shown in Fig.~\ref{fig:delay_0}. In the time delay range where EUV
and laser overlap and in the energy range between $2s^2$ and $2p^2$, the interference pattern
forms fringes along the energy axis, which oscillate in the time delay in the period of the laser
optical cycle (1.8~fs). When the EUV is at the zeros of the laser field (e.g. $t_0=0.9$~fs), the
spectrum is slightly deviated from the $2s2p$ line shape, and the bright fringes appear. When the
EUV is on top of the local maximum of the laser field (e.g., $t_0=1.8$~fs), the $P$-state wave
packet excited by the EUV is distorted, and a large portion of photoelectrons are transferred to
the $S$ and $D$ states. The final wave packet in momentum is less directional in $z$, which
results in the dark fringes. As stated in Sec.~\ref{sec:model}, streaking effect is based on the
linear momentum shift of the electrons and should be described by high angular momentum states in
the spherical coordinates. Since the present model considers the angular momentum only up to
$L=2$, it is intrinsically incapable of quantitatively describing the streaking effect.

\begin{figure}[htbp]
\vspace{0.3cm}
\centering
\includegraphics[width=0.50\textwidth]{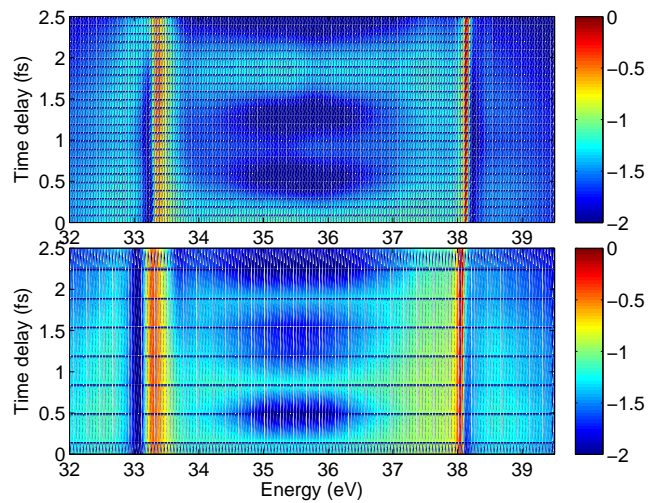}
\caption{(Color online) Transient electron spectra for $t_0=0$--2.5~fs of the $s$ partial waves by
the TAE-TDSE calculations (top panel) and the present model (bottom panel), in logarithmic scale
and in 20-dB dynamic range. The strength of the middle flat part of the spectrum oscillates with
$t_0$ in the period of half optical cycle and is weakened as the laser pulse lags behind the EUV
pulse, but the resonance parts sustain over large time delays.}
\label{fig:delay_S}
\end{figure}

To further illuminate the atomic response near $t_0=0$ in terms of coupled partial waves, we show
the $\epsilon s$ spectra calculated by TAE-TDSE and by the multilevel model for small time delays
in Fig.~\ref{fig:delay_S}. The results agree well between the calculations. A clear feature seen in
the figure is that the middle flat part of the spectrum oscillates alone $t_0$ with roughly half
optical cycles (0.9~fs). In Fig.~\ref{fig:delay_0}, with the alternating phase, $\epsilon s$ is
coherently added to $\epsilon p$ to give the interference pattern oscillating in full optical
cycles. On the contrary, the $\epsilon s$ spectra at the $2s^2$ and $2p^2$ resonances remain
stable over $t_0$. The latter features come from the EUV excitation of $2s2p$ and the
subsequent laser transition to $2s^2$ and $2p^2$ before $2s2p$ decays. This contrast will be
further investigated in view of ultrafast electron dynamics in Sec.~\ref{sec:c-c}.

The sideband-like signals at $2s^2$ and $2p^2$ extend in the time delay up to $t_0=15$~fs.
The formation of the sideband pattern corresponds to two ionization pathways--the direct EUV
ionization to $1s\epsilon p$, and the EUV+VIS ionization through $2s2p$. Thus, the sidebands at
both $2s^2$ and $2p^2$ vanish
as $t_0$ approaches the $2s2p$ decay lifetime (17~fs), where the second ionization pathway
effectively stops. The slopes of the tilted fringes on both resonances change along $t_0$, which
are controlled by the phase differences between $2s2p$ and $2s^2$ and between $2s2p$ and $2p^2$,
respectively, in the duration after the EUV excitation and before the VIS kicks in. This feature
of tilted fringes has been explained by several attosecond ionization studies~\cite{choi,
mauritsson, kim}.

\subsection{Short-time behavior of C-C coupling}\label{sec:c-c}

At $t_0=0$, the effective bandwidth of the laser that corresponds to the dynamics of the wave
packet initiated by the EUV pump is 0.9~eV. Considering this laser bandwidth, the signals in
34-37~eV of the $\epsilon s$ spectrum, shown in Figs.~\ref{fig:xl} and \ref{fig:delay_S},
are only from the laser coupling with the $1s\epsilon p$ background continuum instead of with the
$2s2p$ state. While the C-C dipole matrix elements are fixed values with the atomic structure, the
signals in 34-37~eV vary closely with the laser field in $t_0$, as shown in
Fig.~\ref{fig:delay_S}, i.e., when the EUV pulse is on top of the local peaks of the laser field,
the signals are strong, and when the EUV pulse is at the zero field of the laser, the signals
attenuate. Such behavior suggests that the C-C transition is only significant at the beginning of
the EUV ionization when the electrons are still at the neighborhood of the atomic core. On the
other hand, the excitation of $2s2p$ lasts long after the EUV pulse, and thus the B-B coupling
between $2s2p$ and $2s^2$ or $2p^2$ is insensitive to the time delay between the laser and EUV
pulses.

\begin{figure}[htbp]
\centering
\includegraphics[width=0.50\textwidth]{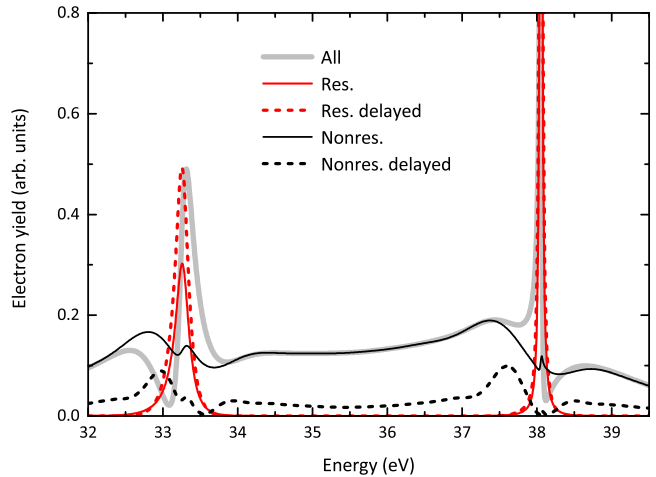}
\caption{(Color online) electron spectra of the $s$ partial wave for three sets of dipole matrix
elements by the model calculation. The first set includes all the dipole matrix elements, while
the second and the third sets include just the nonresonant (C-C) and just the resonant
dipole matrix elements (B-B and B-C), respectively. The peak laser intensity is fixed at
1~TW/cm$^2$. The EUV either overlaps the laser peak or leads in time by 3~fs.}
\label{fig:dip}
\end{figure}

To elucidate the role of the C-C coupling responsible for signals in 34-37~eV in the $\epsilon s$
and $\epsilon d$ spectra, beside the full model calculation with all the original retrieved dipole
matrix elements (described in Sec.~\ref{sec:ml-tdse}), we made two extra calculations with only
the nonresonant (C-C) and only the resonant (B-B and B-C) ones, respectively, for $t_0=0$ and
3~fs. The $t_0=3$~fs case is for non-overlapping pulses where the laser comes later than the EUV.
The resultant $\epsilon s$ electron spectra are plotted in Fig.~\ref{fig:dip}. In the
nonresonant, $t_0=0$ case, the spectrum reproduces the 34--37~eV part of the full calculation
very well. At $2s^2$ and $2p^2$, the resonance peaks shrink to two slight bumps, which are the
results of the configuration interaction to the continuum. When the laser is delayed to
$t_0=3$~fs, the overall nonresonant signals drop dramatically, where the bumps at 33 and 37.6~eV
indicate the ``actual bandwidth'' 0.5~eV of the laser. The effect of the C-C coupling is
substantial only if the two pulses overlap, or more specifically, the EUV overlaps the local
maximum of the laser field, where the EUV-excited continuum electrons can be driven by the laser.
On the contrary, in the resonant case, the flat part
remains 0, and the $2s^2$ and $2p^2$ peaks retain. The bandwidth of the laser slightly changes the
line widths of the resonances, but it has no effect on the empty spectral range in 34--37~eV.

The short-time behavior of the photoelectrons described above can be alternatively explained and
reaffirmed by analyzing the motion of the electron wave packet in the coordinate space. Determined
by the EUV, the wave packet has the central kinetic energy around $2s2p$ at 35.6~eV, corresponding
to the velocity of 67~a.u./fs and the momentum of $k_0=1.6$~a.u. For photoelectrons along, the
asymptotic condition of the continuum wave is determined by $kr \gg 1$, and the ``core'' size of
the system can be defined by $1/k_0=0.6$~a.u. For bound electrons, the core size can be defined
by the bound orbitals, which are less than 1~a.u. for $n \leq 2$. Within the core range, the
emitted electrons still have the chance to interact with each other or with bound electrons.
Considering the core size and the average velocity, it takes less than 20~as for the wave packet
to leave the core, which is significantly shorter than the 0.9~fs half optical cycle of the laser.
In other words, the photoelectrons right after the EUV ionization are bound-like only instantly in
terms of a laser optical cycle before they dissipate away. As a result, in Fig.~\ref{fig:delay_S},
the 34--37~eV signals follow the instantaneous intensity of the laser closely. Meanwhile the
excited $2s2p$ state holds electrons in the timescale of tens of femtoseconds where the laser
couples $2s2p$ to the surrounding of $2s^2$ and $2p^2$, e.g. at 37.5~eV, which shows the slow
attenuation of signals along $t_0$.

The above demonstration shows that the C-C transition in the coupled autoionizing systems is
nonnegligible even in moderately intense fields, in a short period of time after the ionizing
pulse. This is in contrast to the common assumption in previous studies to neglect C-C transitions
when applying long pulses~\cite{lambropoulos, bachau, madsen, themelis} and ultrashort
pulses~\cite{wang, pfeiffer, chu13}. By fitting the model to the TAE-TDSE result, the C-C
dipole matrix elements $D_{1s\epsilon p,1s\epsilon s}$ and $D_{1s\epsilon p,1s\epsilon d}$
are determined to be about 1~a.u./eV. This quantity is equivalent to a 0.1-a.u. dipole matrix
element between the
discrete states of the same widths of $2s2p$ and $2s^2$, which is 15 times smaller $D_{2s2p,2s^2}$.

\section{Conclusions}\label{sec:conclusions}

A general technique has been developed and tested to retrieve the properties of the DFRs in the
autoionization region. An IAP with a synchronized short dressing pulse are applied to an atomic
system to obtain the photoelectron spectra. The simulations are carried out by a multilevel
model and a TAE-TDSE program, where their comparison characterizes the DFRs, in particular the
dipole matrix elements of the transitions involving DFRs. Ultrafast technology is critical in this
technique for the broadband excitation to cover the resonance region in concern. The multilevel
model successfully recovers the electron wave packet for laser intensities up to 1~TW/cm$^2$.
Surprisingly, it shows that the C-C transition by the laser coupling contributes significantly in
the photoelectron
spectra when the laser overlaps the EUV pulse. It indicates the correlation effect between the
``free'' photoelectrons before they move away from the core, in comparison with the bound-state
coupling that has been well studied in laser-dressed systems. The effect depends on the proximity
between the electrons and is controlled by the timing of the laser pulse. The general technique
developed here would help the identification and characterization of DFRs for applications
involving multiphoton processes, and would utilize the most advanced ultrafast optical technologies
in atomic and molecular studies. An algorithm to automatically optimize the DFR parameters could be
developed in the future.

\begin{acknowledgments}
This work was supported in part by Chemical Sciences, Geosciences, and Biosciences Division, Office
of Basic Energy Sciences, Office of Science, US Department of Energy. TM was supported by
Grants-in-Aid for Scientific Research (A), (B), and (C) from the Ministry of Education, Culture,
Sports, Science and Technology, Japan.
\end{acknowledgments}


\end{document}